\documentclass[sigconf]{acmart}

\copyrightyear{2026}
\acmYear{2026}
\acmConference[ICMR '26]{ACM International Conference on Multimedia Retrieval}{June 2026}{Dublin, Ireland}

\usepackage{amsmath}

\usepackage{amssymb}
\usepackage{booktabs}
\usepackage{multirow}
\usepackage{graphicx}
\usepackage{xcolor}

\begin{document}

\title{ReinPool: Reinforcement Learning Pooling Multi-Vector Embeddings for Retrieval System}

\author{Sungguk Cha}
\affiliation{%
  \institution{LG Uplus}
  \city{Seoul}
  \country{South Korea}}
\email{sungguk@lguplus.co.kr}

\author{DongWook Kim}
\affiliation{%
  \institution{LG Uplus}
  \city{Seoul}
  \country{South Korea}}
\email{dongwook92@lguplus.co.kr}

\author{Mintae Kim}
\affiliation{%
  \institution{LG Uplus}
  \city{Seoul}
  \country{South Korea}}
\email{iammt@lguplus.co.kr}

\author{Youngsub Han}
\affiliation{%
  \institution{LG Uplus}
  \city{Seoul}
  \country{South Korea}}
\email{yshan042@lguplus.co.kr}

\author{Byoung-Ki Jeon}
\affiliation{%
  \institution{LG Uplus}
  \city{Seoul}
  \country{South Korea}}
\email{bkjeon@lguplus.co.kr}

\author{Sangyeob Lee}
\affiliation{%
  \institution{LG Uplus}
  \city{Seoul}
  \country{South Korea}}
\email{sangyeob@lguplus.co.kr}

\renewcommand{\shortauthors}{Cha et al.}

\begin{abstract}
Multi-vector embedding models have emerged as a powerful paradigm for document retrieval, preserving fine-grained visual and textual details through token-level representations.
However, this expressiveness comes at a staggering cost: storing embeddings for every token inflates index sizes by over $1000\times$ compared to single-vector approaches, severely limiting scalability.
We introduce \textbf{ReinPool}, a reinforcement learning framework that learns to dynamically filter and pool multi-vector embeddings into compact, retrieval-optimized representations.
By training with an inverse retrieval objective and NDCG-based rewards, ReinPool identifies and retains only the most discriminative vectors without requiring manual importance annotations.
On the Vidore V2 benchmark across three vision-language embedding models, ReinPool compresses multi-vector representations by $746$--$1249\times$ into single vectors while recovering 76--81\% of full multi-vector retrieval performance.
Compared to static mean pooling baselines, ReinPool achieves 22--33\% absolute NDCG@3 improvement, demonstrating that learned selection significantly outperforms heuristic aggregation.
\end{abstract}




\maketitle

\section{Introduction}
Information retrieval is a cornerstone of modern AI systems, enabling applications from search engines to retrieval-augmented generation (RAG) pipelines~\cite{lewis2020rag}.
As these systems evolve toward multimodal capabilities, they must move beyond text-only analysis to understand diverse document formats—slide decks, technical reports, and visually rich PDFs—where layout and visual elements carry significant semantic meaning.

Recent multi-vector embedding models~\cite{faysse2024colpaliefficientdocumentretrieval, huang2025beyond} address this challenge by representing documents as sets of token- or patch-level embeddings, preserving fine-grained visual and textual details through late interaction mechanisms~\cite{khattab2020colbert}.
However, this expressiveness comes at a prohibitive cost: a single document may require over 1,000 embedding vectors, inflating storage and search complexity by orders of magnitude compared to traditional single-vector approaches~\cite{qwen3embedding, johnson2019faiss}.
For instance, the \textit{Tomoro-ColQwen-4b} model produces an average of $1249 \times 320$ dimensional embeddings per document—a $1249\times$ increase over single-vector storage.

\begin{figure}[t]
    \centering
    \includegraphics[width=\columnwidth]{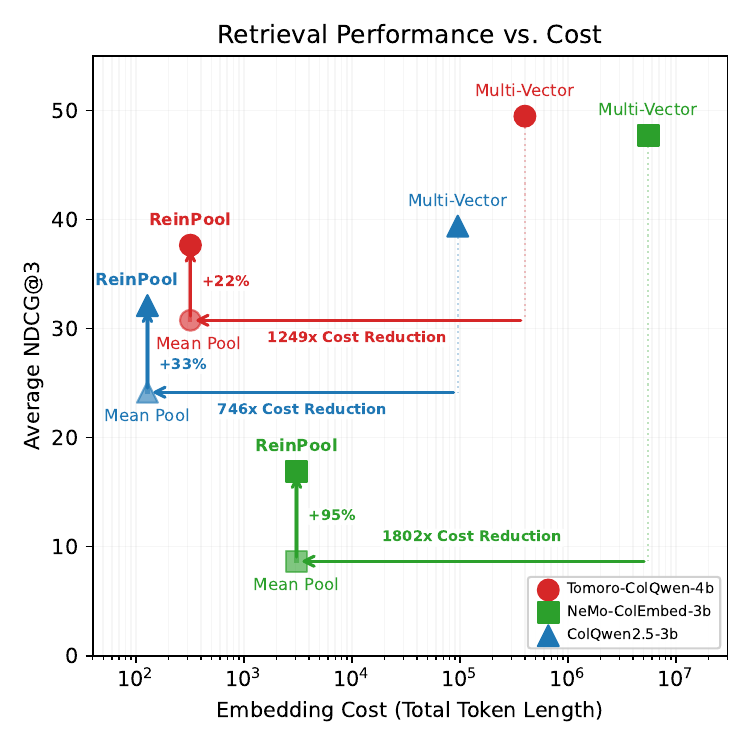}
    \caption{\textbf{Retrieval Performance vs. Cost.} Comparison of ReinPool against Full Multi-Vector and Static Mean Pooling baselines. The x-axis represents the total embedding cost (token length), and the y-axis shows the average NDCG@3 on Vidore V2. Horizontal arrows indicate the massive reduction in retrieval cost; vertical arrows depict the performance gain of ReinPool over static mean pooling.}
    \Description{Scatter plot showing ReinPool achieving better NDCG@3 scores than static pooling at the same embedding cost.}
    \label{fig:efficiency}
\end{figure}

The fundamental inefficiency lies in a lack of selectivity: not all tokens are equally informative for retrieval.
Many embeddings correspond to whitespace, decorative elements, or generic content that dilutes the discriminative signal.
Existing approaches occupy two extremes: storing all vectors preserves detail but is computationally prohibitive, while static pooling (e.g., mean or max) collapses the rich multi-vector representation into a single vector, destroying the fine-grained information that enables precise matching.

We propose \textbf{ReinPool}, a reinforcement learning framework that learns to dynamically filter and pool multi-vector embeddings.
Rather than applying fixed heuristics, ReinPool trains a lightweight policy network to identify and retain only the most retrieval-relevant vectors before aggregation.
The key insight is to frame vector selection as a sequential decision problem, optimizing directly for retrieval performance via an inverse retrieval objective with NDCG-based rewards~\cite{williams1992reinforce}.

Our contributions are:
\begin{itemize}
    \item A novel RL-based pooling framework that optimizes vector selection directly for retrieval effectiveness, rather than reconstruction objectives.
    \item A self-supervised training strategy using synthetic queries and inverse retrieval, eliminating the need for manual importance annotations.
    \item Extensive experiments on Vidore V2 demonstrating that ReinPool achieves $1249\times$ compression while recovering 76\% of full multi-vector performance (37.65 vs. 49.48 NDCG@3), outperforming static mean pooling by 22\% absolute.
\end{itemize}

\section{Related Works}
\label{sec:related_work}

\subsection{Neural Information Retrieval}
The field of neural information retrieval has evolved significantly from traditional lexical methods like BM25~\cite{robertson2009bm25} to neural dense retrieval models.
The introduction of BERT~\cite{devlin2019bert} and the Transformer architecture~\cite{vaswani2017attention} enabled powerful contextualized representations for retrieval.
Early works like Sentence-BERT~\cite{reimers2019sentence} established efficient sentence embeddings using siamese networks, while Dense Passage Retrieval (DPR)~\cite{karpukhin2020dense} demonstrated strong performance on open-domain question answering by encoding queries and documents into single vectors for Maximum Inner Product Search (MIPS).
Contriever~\cite{izacard2022contriever} further advanced unsupervised dense retrieval through contrastive pre-training, achieving competitive performance without labeled data.
However, compressing an entire document into a single vector often results in an information bottleneck, leading to the loss of fine-grained details.

To address this, late interaction models like ColBERT~\cite{khattab2020colbert} proposed representing documents as bags of token embeddings, delaying the interaction until query time to preserve term-level nuances.
ColBERTv2~\cite{santhanam2022colbertv2} improved upon this with residual compression and denoised supervision, achieving state-of-the-art retrieval while reducing index size.
Scaling efforts such as GTR~\cite{ni2022gtr} demonstrated that large dual encoders can be effective generalizable retrievers across diverse domains.
More recently, this paradigm has extended to multimodal domains; models like ColPali~\cite{faysse2024colpaliefficientdocumentretrieval} and Tomoro ColQwen3~\cite{huang2025beyond} leverage Vision Language Models (VLMs) to embed document images directly, capturing visual layout and textual content simultaneously.
CLIP~\cite{radford2021clip} has also influenced multimodal retrieval by learning aligned visual-textual representations.

\subsection{Efficient Indexing for Multi-Vector Models}
While multi-vector models offer superior accuracy, they incur substantial storage and computational costs.
ColBERT, for instance, increases index size by orders of magnitude compared to single-vector models.
Efficient similarity search systems like FAISS~\cite{johnson2019faiss} are essential for scaling retrieval to large corpora.
Approaches like PLAID~\cite{santhanam2022plaid} mitigate the multi-vector overhead by employing aggressive quantization and centroid-based pruning to accelerate retrieval.
BGE~\cite{xiao2024bge} and similar embedding models have explored efficient representations for multilingual and domain-specific retrieval.
Despite these optimizations, the fundamental issue remains: storing embeddings for every token—including non-informative ones like stopwords or background image patches—is inherently inefficient.
ReinPool complements these systems by learning to discard such redundant vectors \textit{before} indexing or quantization, offering a complementary reduction in storage footprint.

\subsection{Learned Pooling and Selection}
Traditional pooling methods (e.g., mean or max pooling) are static and agnostic to downstream retrieval tasks.
Recent works~\cite{gunther2025jinaembeddings, kusupati2024matryoshka} have explored learned pooling mechanisms to bridge the gap between fixed-size and multi-vector representations.
Matryoshka Representation Learning~\cite{kusupati2024matryoshka} enables flexible dimensionality by training nested representations, allowing trade-offs between efficiency and expressiveness at inference time.
However, most existing methods rely on reconstruction objectives or simple attention mechanisms that may not align with retrieval performance.
Our work introduces a reinforcement learning-based approach, optimizing the selection and pooling process directly for retrieval effectiveness (NDCG) via an inverse retrieval formulation.

\section{Methodology}
\label{sec:methodology}

We propose \textbf{ReinPool}, a framework that learns to dynamically filter and pool multi-vector embeddings.
In this section, we first formulate the document representation problem, then introduce the ReinPool agent, and finally describe the training strategy based on inverse retrieval.

\subsection{Problem Formulation}
\label{sec:problem}
Let a document $d$ be projected into a sequence of $N$ embedding vectors $V_d = \{v_1, v_2, \dots, v_N\}$, where $v_i \in \mathbb{R}^{d_{model}}$. 
While storing $V_d$ fully preserves fine-grained details, it incurs high storage and retrieval costs. 
Our goal is to learn a mapping $f: \mathbb{R}^{N \times d_{model}} \rightarrow \mathbb{R}^{d_{model}}$ that compresses the multi-vector sequence into a single vector $\mathbf{v}_{\text{pool}}$, while maximizing retrieval performance.
ReinPool achieves this by selectively filtering vectors via a reinforcement learning policy before aggregating the retained set.

\subsection{ReinPool Framework}
\label{sec:reinpool}
The ReinPool process consists of two stages: \textbf{Filtering} and \textbf{Pooling}.

\subsubsection{Filtering Policy}
We employ a policy network $\pi_\theta$ to identify a subset of informative vectors.
For each vector $v_t \in V_d$, the agent outputs a binary action $a_t \in \{0, 1\}$:
\begin{itemize}
    \item If $a_t = 1$: Keep vector $v_t$.
    \item If $a_t = 0$: Discard vector $v_t$.
\end{itemize}
The set of retained vectors is denoted as $V_{\text{keep}} = \{v_t \mid a_t = 1\}$.

The policy model $\pi_\theta$ is implemented as a Transformer-based architecture~\cite{vaswani2017attention} comprising a multi-head attention layer (8 heads) followed by a linear classification head.
Crucially, the number of parameters in $\pi_\theta$ is $O(d_{model}^2)$, which is negligible compared to the underlying embedding model.
The objective of $\pi_\theta$ is not to transform the embedding space, but rather to eliminate redundant or non-informative vectors.

\subsubsection{Pooling}
The filtered subset of vectors is aggregated into a single representation $\mathbf{v}_{\text{pool}} \in \mathbb{R}^{d_{model}}$ using a standard pooling function (e.g., Mean or Max pooling):
\begin{equation}
    \mathbf{v}_{\text{pool}} = \text{Pool}(V_{\text{keep}})
\end{equation}
This results in a standard dense embedding that fits into existing single-vector search indices.

\subsection{Training Strategy}
\label{sec:training}
To effectively train the filtering policy $\pi_\theta$ without requiring annotated relevance labels, we propose a self-supervised training framework based on \textit{inverse retrieval}.
Unlike standard retrieval where a query is used to search for documents, our training objective inverts this relationship: the agent learns to pool a document such that the resulting vector provides high similarity to its corresponding synthetic queries.

\subsubsection{Synthetic Query Generation}
We construct a training dataset by generating synthetic queries for each document in our corpus $\mathcal{D}$.
For each document $d \in \mathcal{D}$, we utilize a Large Language Model (LLM)~\cite{ouyang2022instructgpt} to generate a set of $k$ distinct queries $\mathcal{Q}_d = \{q_1, \dots, q_k\}$ relevant to $d$.
This process yields a set of positive pairs $(d, q^+)$ where $q^+ \in \mathcal{Q}_d$.

\subsubsection{Pre-computation and Setup}
To enable efficient RL training, we treat the underlying embedding space as fixed.
We pre-compute the multi-vector embeddings for all documents and the single-vector embeddings for all generated queries using the base embedding model $\mathcal{M}$:
\begin{align}
    V_d &= \mathcal{M}(d) \in \mathbb{R}^{N \times d_{model}} \\
    V_q &= \mathcal{M}(q) \in \mathbb{R}^{N \times d_{model}}
\end{align}
These pre-computed embeddings serve as the environment state for the RL agent.

\subsubsection{Optimization via Inverse Retrieval Rank}
The training process iterates through the following steps to optimize the policy $\pi_\theta$:

\begin{enumerate}
    \item \textbf{Policy Rollout:} Given a document embedding $V_d$, the policy samples a mask $a \sim \pi_\theta(\cdot|V_d)$ to select a subset of vectors $V_{\text{keep}}$.
    
    \item \textbf{Pooling:} The retained vectors are aggregated into a single document representation:
    \begin{equation}
        \mathbf{v}_{\text{pool}} = \text{Pool}(V_{\text{keep}})
    \end{equation}
    
    \item \textbf{Inverse Retrieval Simulation:} We treat $\mathbf{v}_{\text{pool}}$ as a query vector to search against the synthetic queries $\mathcal{Q}$.
    The objective is to retrieve the ground-truth query $V_{q^+}$ associated with document $d$ from $\mathcal{Q}$.
    Similarity scores are computed between $\mathbf{v}_{\text{pool}}$ and all queries in $\mathcal{Q}$:
    \begin{equation}
        s(d, \mathcal{Q}) = \mathbf{v}_{\text{pool}}^\top \text{Pool}(V_{q})
    \end{equation}
    
    \item \textbf{Rank-Based Reward:} 
    We align the training objective with retrieval performance by using Normalized Discounted Cumulative Gain (NDCG)~\cite{jarvelin2002cumulated} as the reward signal. 
    NDCG is a standard information retrieval metric that evaluating ranking quality not only by the presence of relevant items, but also by their position in the list.
    For a given document $d$ and its ground-truth query $q^+$, we rank the set of synthetic queries $\mathcal{Q}$ based on their computed similarity scores.
    The reward is defined as:
    \begin{equation}
        r = \text{NDCG}(q^+, \mathcal{Q})
    \end{equation}
    This formulation applies a logarithmic discount to the rank of the positive query, effectively penalizing the agent more for missing the top spots than for ordering errors deeper in the list. 
    This encourages the model to prioritize features that are highly discriminative for the specific query.
    
    \item \textbf{Policy Update:} The policy parameters are updated using Group Relative Policy Optimization (GRPO)~\cite{shao2024grpo} to maximize the expected reward.
    GRPO estimates advantages by centering rewards within a group of samples, eliminating the need for a separate value network while maintaining stable policy updates similar to PPO~\cite{schulman2017ppo}.
\end{enumerate}

\begin{table*}[t]
    \caption{Information retrieval results on Vidore V2 in NDCG@3. Embedding cost refers to average token length of the embedded corpus.}
    \label{table:quantitative}
    \centering
    \begin{tabular}{lllrrccccc}
        \toprule
        Embedding & Query & Corpus & Embedding & \multicolumn{5}{c}{Vidore V2} \\
        Model & Pooling & Pooling & Cost & ESG & ECO & ESGHL & BIO & AVG \\
        \midrule
        \multirow{6}{*}{tomoro-colqwen-4b} & \multirow{3}{*}{Mean} & \color{gray} - & \color{gray} $1249 \times 320$ & \color{gray} 44.59 & \color{gray} 50.28 & \color{gray} 46.77 & \color{gray} 56.28 & \color{gray} 49.48 \\
        & & Mean & $1 \times 320$ & 16.65 & 50.30 & 13.06 & 43.04 & 30.76 \\
        & & $\text{ReinPool}_{\text{Mean}}$ & $1 \times 320$ & 27.80 & 57.60 & 21.80 & 43.41 & \textbf{37.65} \\
        & \multirow{3}{*}{Max} & \color{gray} - & \color{gray} $1249 \times 320$ & \color{gray} 3.19 & \color{gray} 15.24 & \color{gray} 9.10 & \color{gray} 13.87 & \color{gray} 39.40 \\ 
        & & Max & $1 \times 320$ & 0 & 4.49 & 1.92 & 2.65 & 2.27 \\
        & & $\text{ReinPool}_{\text{Max}}$ & $1 \times 320$ & 4.56 & 20.80 & 4.45 & 3.37 & 8.30 \\
        \midrule
        \multirow{6}{*}{nemo-colembed-3b} & \multirow{3}{*}{Mean} & \color{gray} - & \color{gray} $1802 \times 3072$ & \color{gray} 44.77 & \color{gray} 49.42  & \color{gray} 43.22 & \color{gray} 53.57 & \color{gray} 47.75 \\ 
        & & Mean & $1 \times 3072$ & 3.20 & 0.81 & 11.34 & 19.29 & 8.66 \\
        & & $\text{ReinPool}_{\text{Mean}}$ & $1 \times 3072$ & 14.01 & 18.49 & 13.04  & 22.07  & \textbf{16.90} \\
        & \multirow{3}{*}{Max} & \color{gray} - & \color{gray} $1082 \times 3072$ & \color{gray} 27.17  & \color{gray} 36.40 & \color{gray} 23.85 & \color{gray} 45.86 & \color{gray} 33.32 \\ 
        & & Max & $1 \times 3072$ & 1.36 & 33.68 & 2.32 & 3.72 & 10.27 \\
        & & $\text{ReinPool}_{\text{Max}}$ & $1 \times 3072$ & 5.12 & 34.78 & 5.00 & 3.90 & 12.20 \\
        \midrule
        \multirow{6}{*}{colqwen2.5-3b} & \multirow{3}{*}{Mean} & \color{gray} - & \color{gray} $746 \times 128$ & \color{gray} 34.05 & \color{gray} 49.65 & \color{gray} 30.72 & \color{gray} 43.16 & \color{gray} 39.40 \\ 
        & & Mean & $1 \times 128$ & 13.62 & 49.12 & 11.23 & 23.67 & 24.16 \\
        & & $\text{ReinPool}_{\text{Mean}}$ & $1 \times 128$ & 32.98 & 55.38 & 14.68 & 25.40 & \textbf{32.11} \\
        & \multirow{3}{*}{Max} & \color{gray} - & \color{gray} $746 \times 128$ & \color{gray} 1.04 & \color{gray} 16.36 & \color{gray} 0.59 & \color{gray} 2.21 & \color{gray} 5.05 \\ 
        & & Max & $1 \times 128$ & 1.75 & 1.62 & 0.57 & 0.72 & 1.17 \\
        & & $\text{ReinPool}_{\text{Max}}$ & $1 \times 128$ & 4.64 & 13.28 & 4.63 & 3.46 & 6.50 \\
        \bottomrule
    \end{tabular}
\end{table*}

\section{Experiments}
\label{sec:experiments}

We evaluate ReinPool on the Vidore V2 benchmark to answer three research questions:
(1) What compression ratio does ReinPool achieve versus full multi-vector storage?
(2) How much retrieval performance is preserved after compression?
(3) Does learned pooling outperform static pooling baselines?

\subsection{Experimental Setup}

\textbf{Benchmark.}
We evaluate on \textbf{Vidore V2}~\cite{vidore2024}, a comprehensive benchmark for visual document retrieval spanning four domains: ESG (Environmental, Social, Governance reports), ECO (Economic documents), ESGHL (ESG Highlights), and BIO (Biomedical literature).
We report NDCG@3 for each subset and the average across all tasks.

\textbf{Embedding Models.}
We apply ReinPool to three state-of-the-art vision-language embedding models with varying architectures:
\begin{itemize}
    \item \textbf{Tomoro-ColQwen-4b}~\cite{huang2025beyond}: 4B parameters, hidden dimension 320, average 1249 vectors per document.
    \item \textbf{NeMo-ColEmbed-3b}~\cite{nemoretriever}: 3B parameters, hidden dimension 3072, average 1802 vectors per document.
    \item \textbf{ColQwen2.5-3b}~\cite{colqwen25}: 3B parameters, compact dimension 128, average 746 vectors per document.
\end{itemize}

\textbf{Baselines.}
We compare against:
(1) \textbf{Full Multi-Vector}: the uncompressed representation storing all token embeddings;
(2) \textbf{Static Mean/Max Pooling}: naive aggregation of all vectors into a single representation.

\textbf{Implementation Details.}
We implement ReinPool using the PyTorch framework. 
The filtering policy network is trained using the AdamW optimizer with a weight decay of $0.01$. 
We employ a learning rate scheduler (\texttt{ReduceLROnPlateau}) that halves the learning rate when the validation NDCG@3 score plateaus, ensuring fine-grained convergence. 
Gradients are clipped at a global norm of $1.0$ to stabilize training.
The RL training loop follows the Group Relative Policy Optimization (GRPO)~\cite{shao2024grpo} paradigm. 
For every training step, inputs are repeated to generate a group of variations (determined by the \texttt{group\_size} hyperparameter) by sampling masks from the policy. 
The reward signal is the NDCG@3 score computed against the ground-truth queries associated with the document. 
Advantages are estimated by centering the rewards within each group, effectively normalizing the signal without requiring a separate value network. 
The policy is updated to maximize the expected advantage-weighted log-likelihood of the selected actions. 
The base embedding models remain frozen throughout the training of the ReinPool policy to ensure computational efficiency.

\subsection{Main Results}

Table~\ref{table:quantitative} presents the retrieval results.
We report Embedding Cost as the total dimensionality per document (Number of Vectors $\times$ Vector Dimension).

\textbf{Compression Efficiency.}
ReinPool achieves dramatic compression ratios across all models.
For Tomoro-ColQwen-4b, storage is reduced from $1249 \times 320$ to $1 \times 320$—a \textbf{1249$\times$ reduction}—while maintaining 76\% of retrieval performance (37.65 vs. 49.48 NDCG@3).
Similarly, ColQwen2.5-3b is compressed from $746 \times 128$ to $1 \times 128$ while recovering 81\% of full performance (32.11 vs. 39.40).

\textbf{Superiority over Static Pooling.}
ReinPool consistently and substantially outperforms naive pooling.
On Tomoro-ColQwen-4b, $\text{ReinPool}_{\text{Mean}}$ achieves 37.65 NDCG@3, a \textbf{22\% absolute improvement} over static mean pooling (30.76).
For ColQwen2.5-3b, the improvement is even more pronounced: $\text{ReinPool}_{\text{Mean}}$ achieves 32.11 versus 24.16 for static mean (\textbf{+33\% relative}).
These gains demonstrate that the RL policy successfully identifies and preserves discriminative vectors that static methods dilute with non-informative content.

\textbf{Effect of Embedding Dimension.}
Compression effectiveness correlates with embedding dimensionality.
For NeMo-ColEmbed-3b (dim 3072), both static pooling (8.66) and ReinPool (16.90) struggle to match full multi-vector performance (47.75), suggesting that extremely high-dimensional spaces contain fine-grained information difficult to compress into a single vector.
However, ReinPool still achieves a \textbf{95\% relative improvement} over static mean pooling (16.90 vs. 8.66).
Conversely, lower-dimensional models like ColQwen2.5-3b (dim 128) are highly amenable to learned pooling.

\textbf{Mean vs. Max Pooling.}
$\text{ReinPool}_{\text{Mean}}$ consistently outperforms $\text{ReinPool}_{\text{Max}}$ across all models, suggesting that mean aggregation better preserves the distributional properties learned by the embedding models.
Nevertheless, $\text{ReinPool}_{\text{Max}}$ still provides substantial gains over static max pooling (e.g., 8.30 vs. 2.27 on Tomoro-ColQwen-4b).

\section{Conclusion}

We presented ReinPool, a reinforcement learning framework for compressing multi-vector document embeddings into efficient single-vector representations.
By framing vector selection as a sequential decision problem and training with an inverse retrieval objective, ReinPool learns to identify and retain only the most discriminative embeddings without requiring manual annotations.

Our experiments on Vidore V2 demonstrate that ReinPool achieves compression ratios exceeding $1000\times$ while preserving up to 81\% of full multi-vector retrieval performance.
Compared to static pooling baselines, ReinPool provides consistent improvements of 22--33\% absolute NDCG, validating that learned selection significantly outperforms heuristic aggregation.

These results establish ReinPool as a practical solution for deploying multi-vector retrieval at scale.
By bridging the gap between single-vector efficiency and multi-vector precision, our approach enables document-native retrieval systems that were previously computationally prohibitive.



\end{document}